# A. Photon findings

B. **Author name:** Víctor M. Urbina Bolland

C. **Affiliation:** Química Foliar S.A. de C.V.

Avenida Urbina No. 4, Parque Industrial Naucalpan

Naucalpan, Estado de México. C.P:53489. México

Tel: (52) 5300 3571, Fax: (52) 5301 0863

E-mail: qfoliar@mail.internet.com.mx

http://light.hinet.com.mx

## E. Abstract:

Two experiments were made using a microwave generator, which sent a narrow beam, through a metallic plate with horizontal movement. At the other end a horn antenna coupled to a field-strength detector.

In linear polarization double cycloids paths were found and in circular polarization spiral paths were found.

These experiments suggested that the photon is composed by **two particles** in dynamic equilibrium. The description of this model is given later as well as its parameters.

## 1. PHOTON FINDINGS

### A. Linear polarization experiment:

For this experiment a Gunn diode oscillator was used coupled to a resonant cavity which frequency was adjusted to 10.5 Ghz, 2.855 cm wavelength without modulation, focused by a parabolic antenna.

A horizontal movement system with a metallic plate was installed with the edge at the center of the microwave beam, measuring every millimeter, as seen in FIG. 1

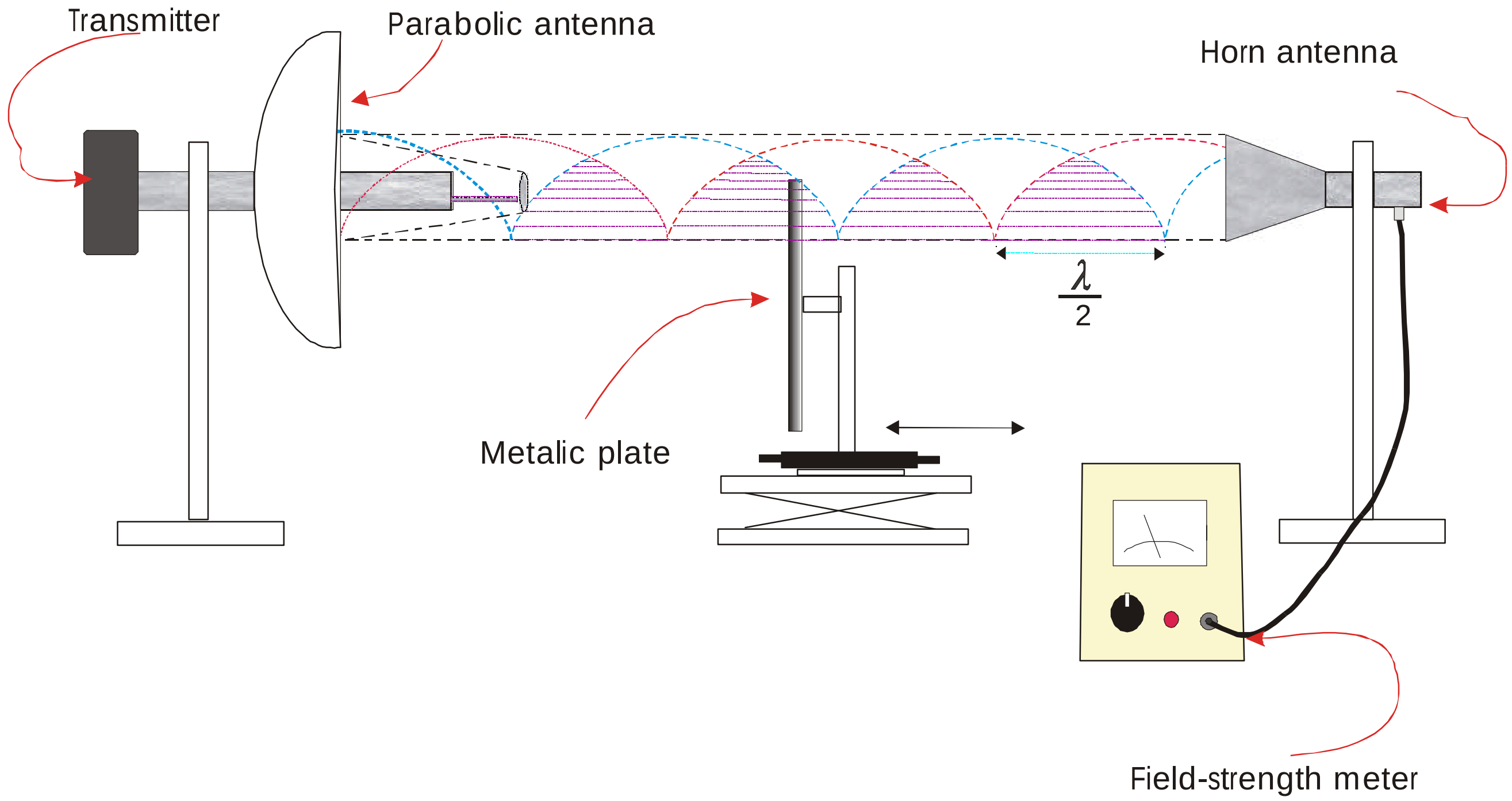

FIG. 1 experiment diagram

Obtaining the following results (Table I):

| Displacement in Cm | Field strength uW | Displacement in Cm | Field strength uW |
|---|---|---|---|
| 10 | 46 | 11.3 | 46 |
| 10.1 | 46 | 11.4 | 47 |
| 10.2 | 43 | 11.5 | 46 |
| 10.3 | 38 | 11.6 | 43 |
| 10.4 | 32 | 11.7 | 39 |
| 10.5 | 19 | 11.8 | 35 |
| 10.6 | 9 | 11.9 | 28 |
| 10.7 | 8 | 12 | 16 |
| 10.8 | 16 | 12.1 | 7 |
| 10.9 | 26 | 12.2 | 8 |
| 11.0 | 35 | 12.3 | 18 |
| 11.1 | 39 | 12.4 | 30 |
| 11.2 | 42 | 12.5 | 36 |
| | | 12.6 | 41 |

Table I: Data obtained

These results were plotted where the horizontal axis coincide with the movement of the plate in centimeters and vertically the field strength in microwatts FIG.2:

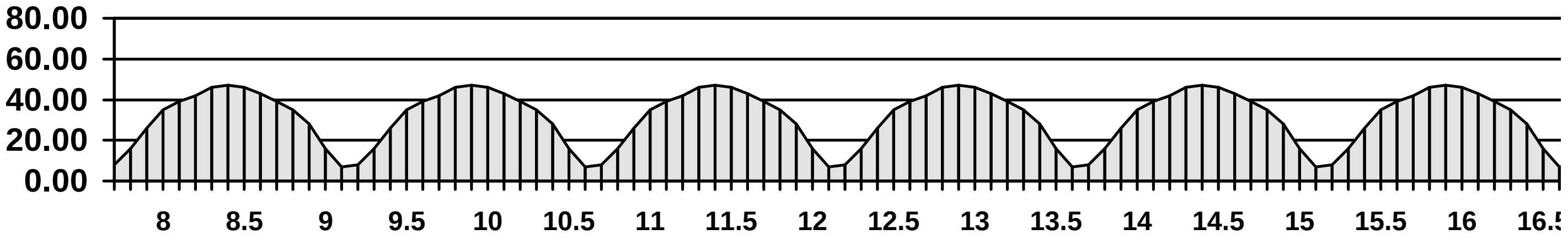

FIG.2 Graphic

In this graphic it is clearly observed that the trajectories are not senoidals but coincides exactly with the empty spaces of the plate following double cycloidal paths FIG.3, which are generated by a spinning wheel, which the tangential speed is the same as its translational speed.

It is interesting to observe that the peaks of minimum power coincide exactly with half of the wavelength, which is 1.4 cm.

It is demonstrated that the path of the electromagnetic radiations is not senoidal. Instead, the trajectories are cycloidal in plane polarization. The simple harmonic movement generates this path.

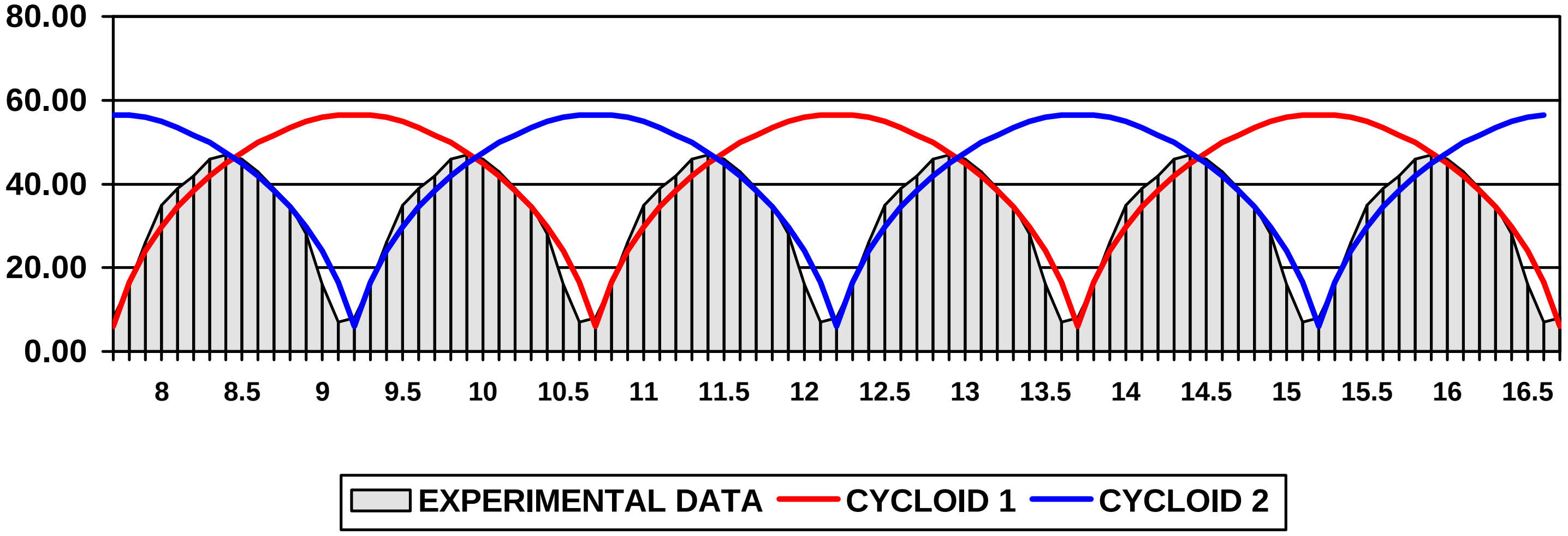

FIG.3 Double cycloid

## B. Circular polarization experiment:

In this experiment the same equipment was used but instead of the parabolic antenna, two helical antennas were used. The edge of the metallic plate at the center of the beam and was moved horizontally every millimeter, measuring the field strength. As shown in FIG.4

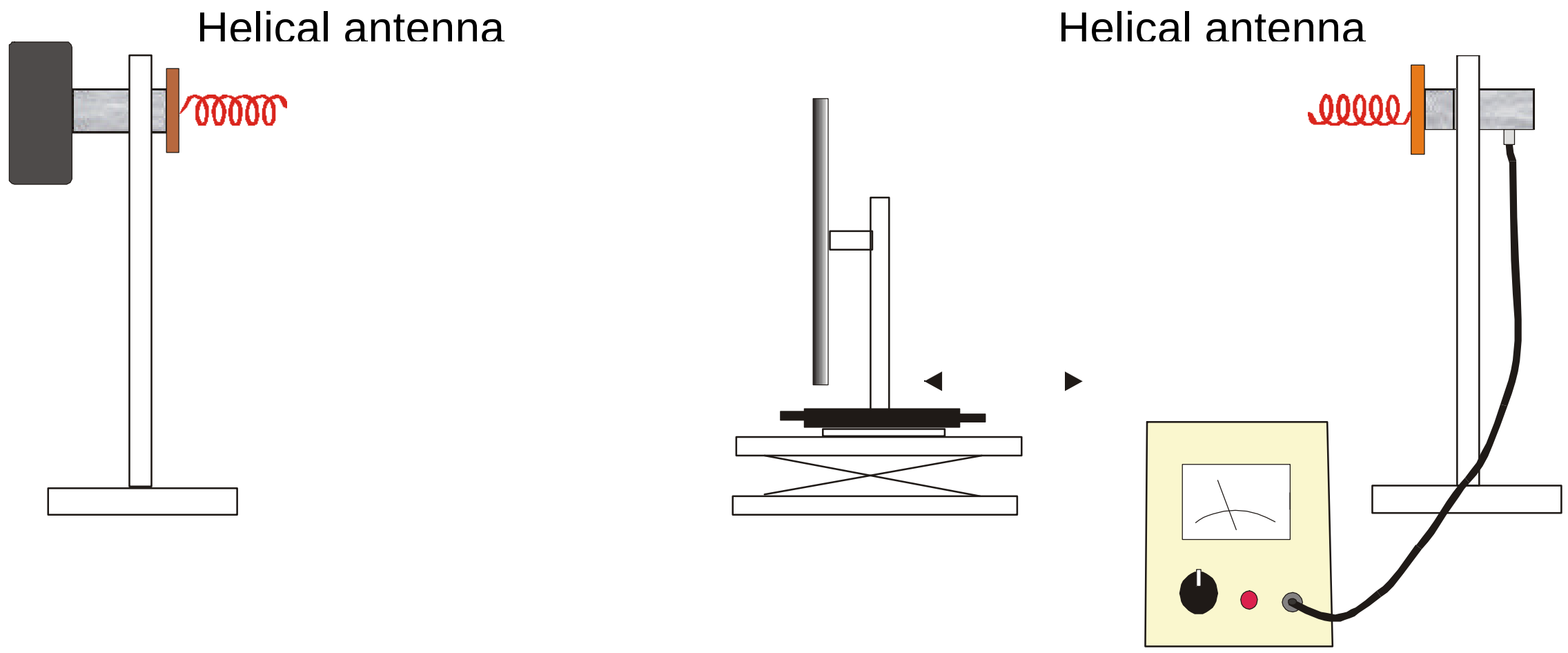

FIG.4 Second experiment diagram

Obtaining the following results (Table II):

| Displacement in mm | Field strength in uW | Displacement in mm | Field strength in uW |
|---|---|---|---|
| 8 | 0 | 21 | 1 |
| 9 | 1 | 22 | 2 |
| 10 | 2.5 | 23 | 2.5 |
| 11 | 4.5 | 24 | 4 |
| 12 | 8 | 25 | 5.5 |
| 13 | 10 | 26 | 7 |
| 14 | 12.5 | 27 | 11 |
| 15 | 13 | 28 | 12.5 |
| 16 | 11.5 | 29 | 13 |
| 17 | 10 | 30 | 11.5 |
| 18 | 7 | 31 | 9 |
| 19 | 4 | 32 | 7.5 |
| 20 | 2 | 33 | 5 |
| | | 34 | 3 |

Table II. Second experiment data obtained

Plotting in X axis distance in millimeters and the filed strength in Y axis (microwatts) FIG.5 was obtained:

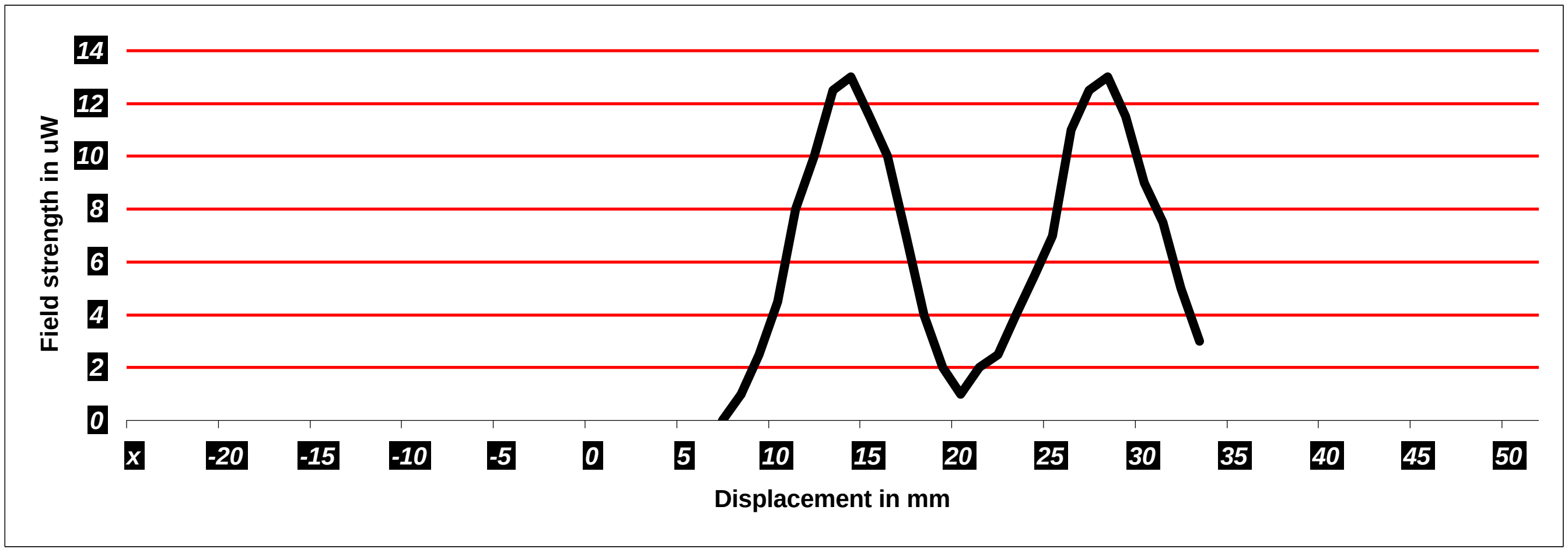

FIG.5 Circular polarization

The two dimensional projection of two spiral trajectories is two senoidal figures and if the previous graph is overlapped, the following figure is obtained. FIG.8

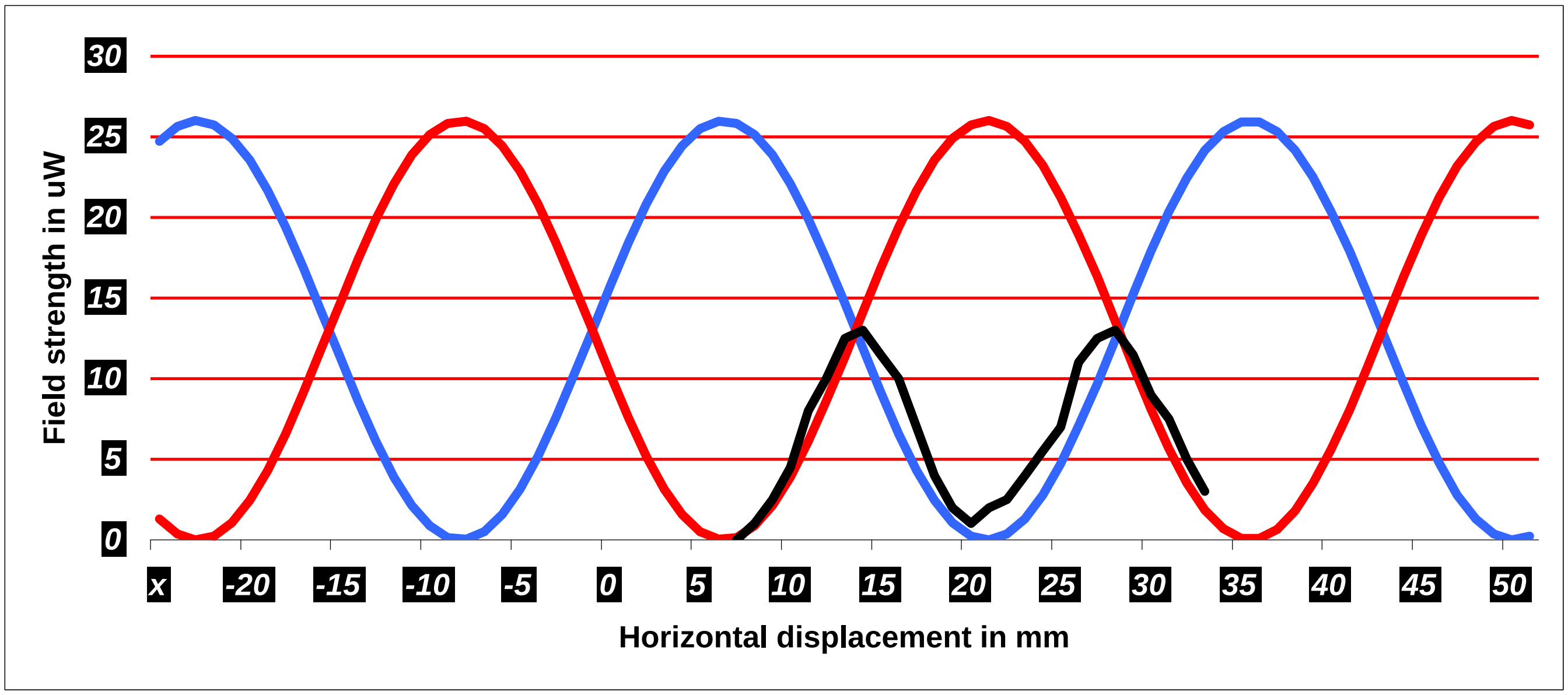

FIG.6 Helical trajectories

These experiments suggest the following photon model:

## C. **Proposed photon model**

Consider, in dynamic equilibrium, a photon formed by two particles whose electric charges are opposed, and for this reason they attract each other. This attraction is equilibrated by the centrifugal force generated when rotating one with the other. (See FIG. 7). By rotating electrical charged particles, a perpendicular magnetic field is generated, also perpendicular to a stationary observer.

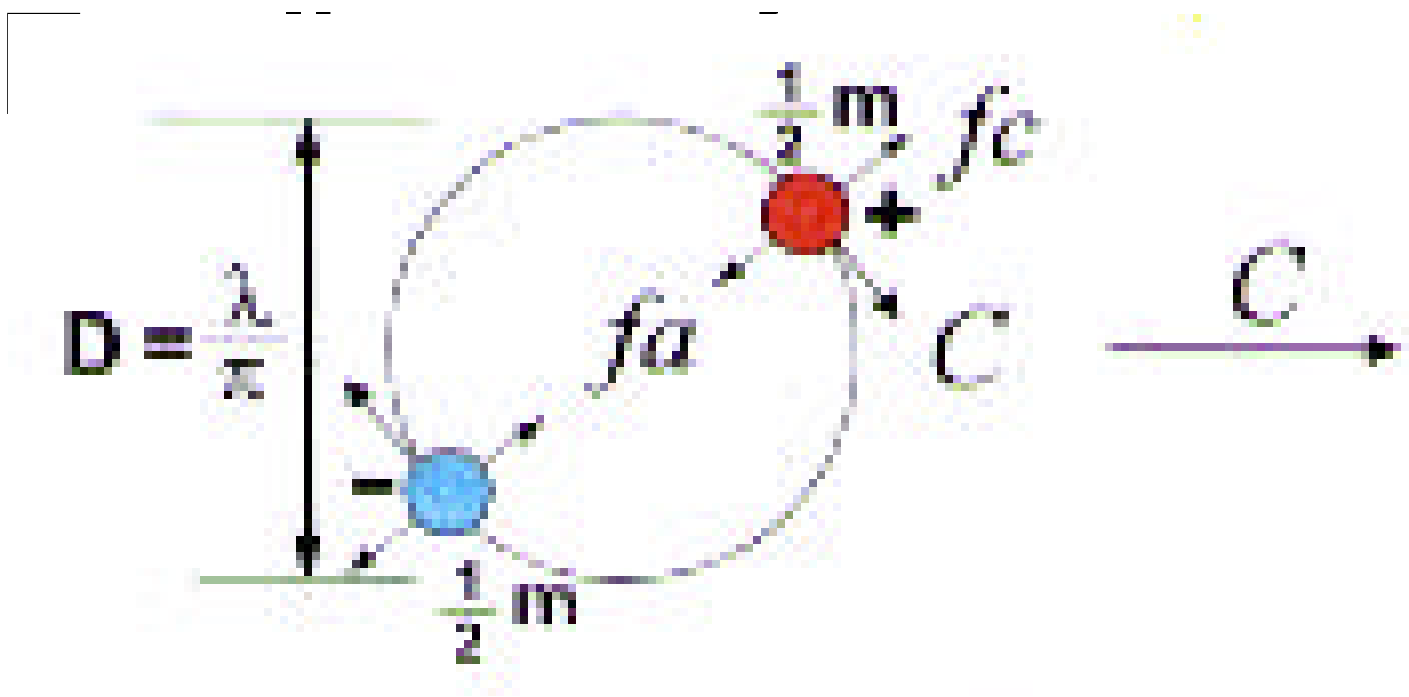

FIG.7 Photon model

Charge $$q = \sqrt{\frac{hc}{pk}} \quad (1)$$

h = Planck Constant ( $6.626 \times 10^{-34} \frac{joules}{sec.}$ )

c = Light velocity in vacuum ( $2.998 \times 10^{8} \frac{m}{s}$ )

k = Coulomb constant ( $8.897 \times 10^{9} \frac{Nm^2}{C^2}$ )

Following are definitions of the considered photon:

### ***1. Distance between particles***:

It is exactly the wavelength divided by $p$ (3.1416):

$$d = \frac{l}{p} \quad (2)$$

**2.** :

C = $(2. \quad 10^{8}$ — *(3)*

***Particle mass***

It is relativistic; in other words, depends on its frequency. Individual particle mass is equally

Individual particle mass: $\frac{}{2} = \frac{hf}{2C^{2}}$ *(4)*

***4. Particle charge***:

It is calculated equaling centrifugal force to the electrostatic attraction force as follows:

$$\frac{mC^{2}}{2r} = k\frac{q^{2}}{4r^{2}} \qquad (5)$$

hence $q^{2} = 2\frac{mc^{2}r}{k}$ *(6)*

Since $m = \frac{hf}{C^{2}}$ *(7)*

Replacing *(7)* in *(6)*: $q^{2} = \frac{2hfr}{k}$ But $C = fl$ *(8)* and $l = 2pr$ *(9)*

Hence $c = 2\,fpr$ therefore $fr = \frac{c}{2p}$

Then $q^{2} = \frac{hC}{pk}$ therefore $q = \sqrt{\frac{hC}{pk}}$

A constant value is obtained, by substituting the values:

$$q = \sqrt{\frac{(6.26\times10^{-34})(2.998\times10^{8})}{3.1459(8.987\times10^{9})}} \qquad (10)$$

$$q = 2.653\times10^{-18}\,Coulombs$$

This charge if 16.558 times an electron charge.

## 5. ***System energy***:

It is the sum of translational kinetic energy plus rotational kinetic energy and it is expressed:

Total energy: $E_{total} = E_{translational-kinetic} + E_{rotacional-kinetic}$

$$E = E_{kt} + E_{kr} \qquad \textit{(11)}$$

$$E_{kt} = \tfrac{1}{2} \times \tfrac{1}{2} \times 2mc^2 = \frac{mc^2}{2} \qquad \textit{(12)}$$

$$E_{kr} = \tfrac{1}{2} \times \tfrac{1}{2} \times 2Iw^2 = \frac{Iw^2}{2} \qquad \textit{(13)}$$

The moment of inertia for a particle body is:

$$I = mr^2 \qquad \textit{(14)}$$

By substituting equation *(14)* into *(13)*:

$$E_{kr} = \frac{mr^2 w^2}{2}$$

And since the particle tangential velocity equals c, then:

$$r^2 w^2 = c^2 \qquad \text{hence} \qquad E_{kr} = \frac{mc^2}{2}$$

By substituting in *(11)*:

$$E = \frac{mc^2}{2} + \frac{mc^2}{2}$$

Which yields the well-known equation:

$$E = mc^2$$

This equation would not be obtained if the photon was a single particle.

### *6. Trajectories*:

It is variable, depending on polarization; if polarization is linear, the gyration plane coincides with the translational displacement, forming cycloidal trajectories.

When the translational movement is perpendicular to the rotation plane, then the trajectory is in spiral, called circular polarization. Any intermediate position will produce an elliptical polarization. (FIG.8).

Linear polarization

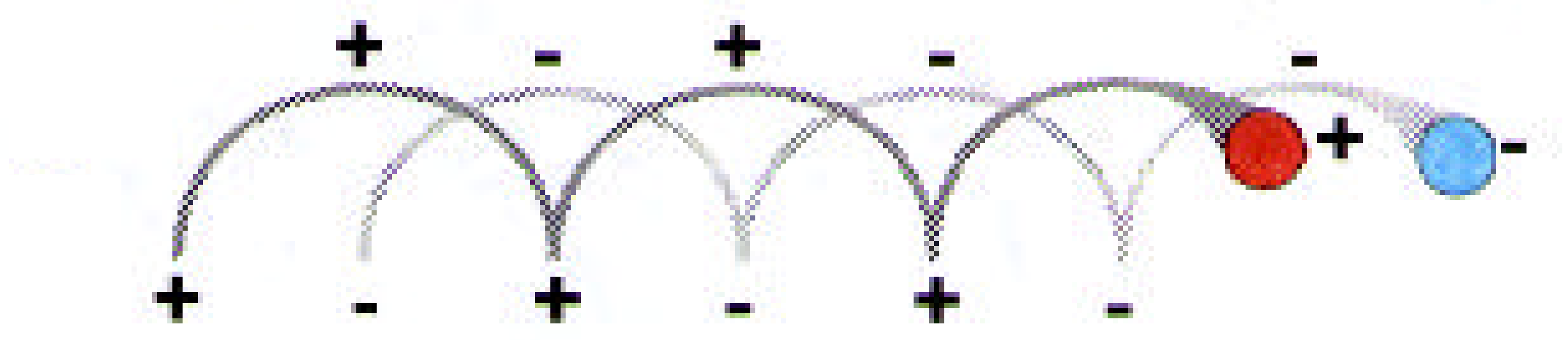

Circular polarization

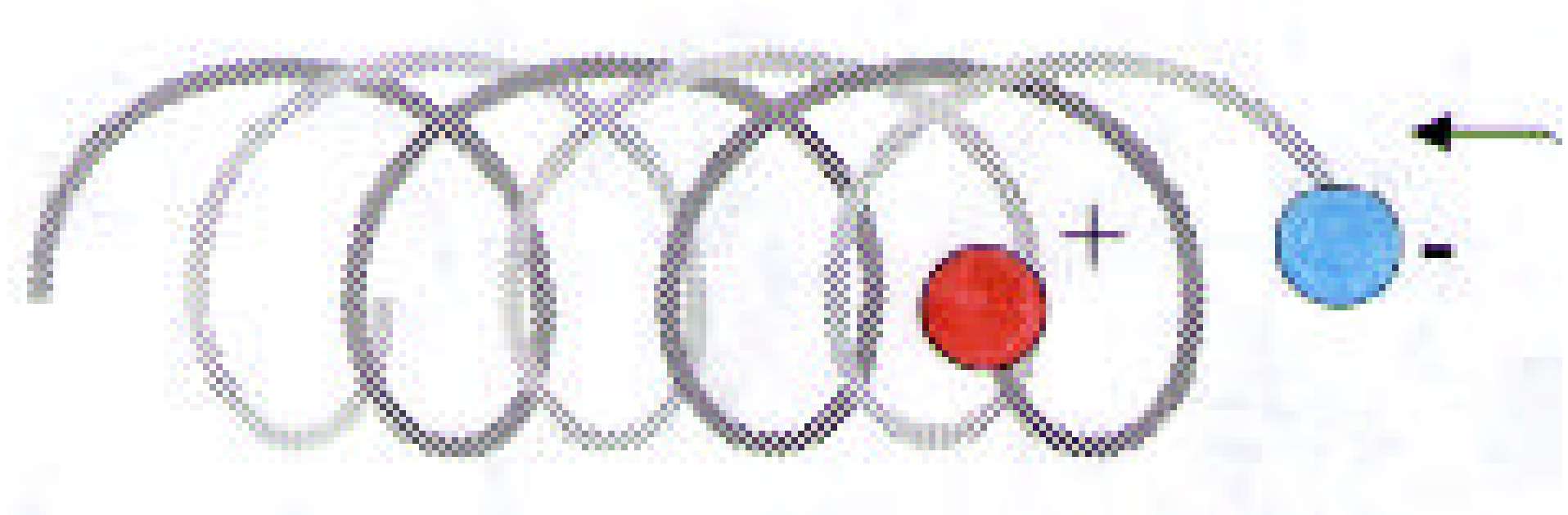

Eliptical polarization

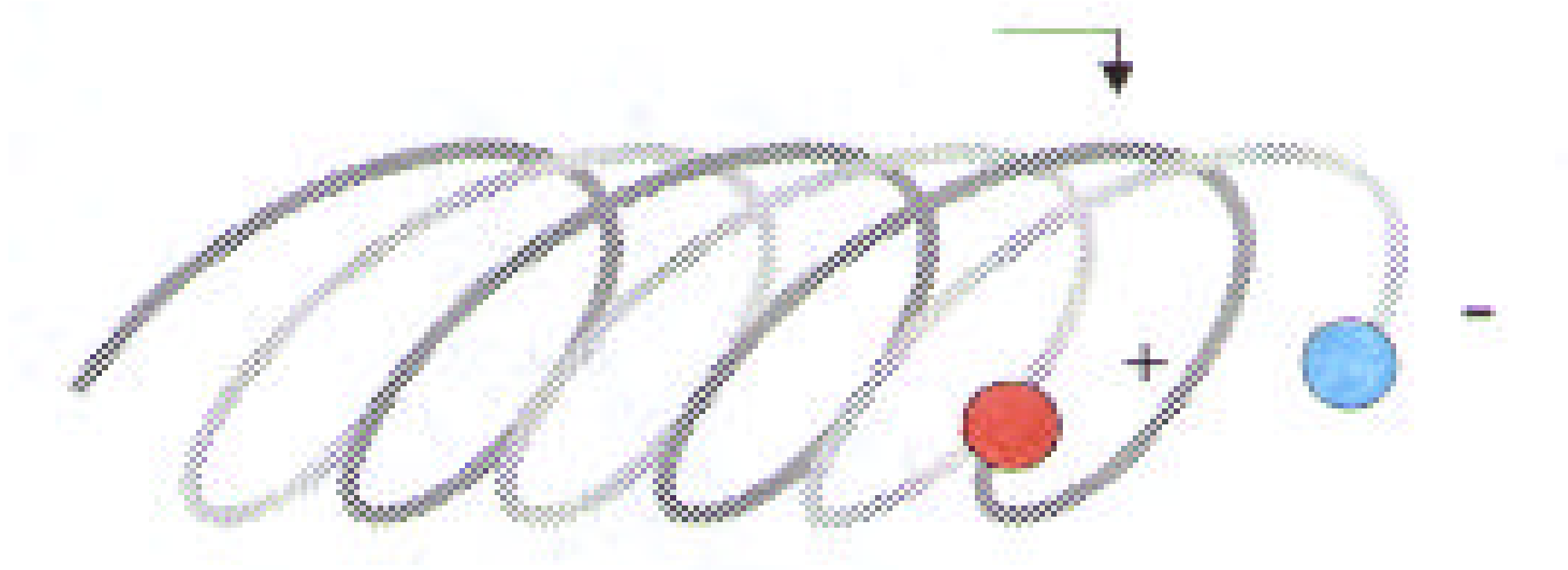

FIG.8 Different trajectories

# D. GLOSSARY

| | |
|---|---|
| m | Photon mass |
| c | light velocity in vacuum ($2.998\times10^{8}\,\frac{m}{s}$) |
| r | particle radius of gyration |
| k | Coulomb constant ($8.897\times10^{9}\,\frac{Nm^2}{C^2}$) |
| f | frequency, Hz. |
| $l$ | wavelength, m. |
| h | Planck constant ($6.626\times10^{-34}\,\frac{joules}{\text{sec.}}$) |
| $p$ | 3.1416 |
| q | particle charge ($2.653\times10^{-18}\,Coulombs$) |
| E | energy in Joules. |
| d | distance between particles (2r) |
| Ekt | translational kinetic energy, *Joules.* |
| Ekr | rotational kinetic energy, *Joules.* |
| I | inertia moment $Kg\cdot m^2$ |
| $w$ | angular velocity, rad / sec. |